\documentclass[12pt]{iopart}

\usepackage{iopams}
\eqnobysec

\usepackage{graphics}
\usepackage{dcolumn}
\usepackage{bm}
\usepackage{hyperref}

\usepackage{amssymb}
\usepackage{amsfonts}

\begin{document}

\title[Optimal basis set]{Optimal basis set for ab-initio calculations of energy levels in tunneling structures, using the covariance matrix of the wave functions}

\author{AS. Spanulescu$^{1}$}

\address{$^{1}$Department of Physics, Hyperion University of Bucharest, Postal code
030629, Bucharest, Romania}


\ead{severspa2004@yahoo.com}
\begin{abstract}
The paper proposes a method to obtain the optimal basis set for
solving the self consistent field (SCF) equations for large atomic
systems in order to calculate the energy barriers in tunneling
structures, with higher accuracy and speed. Taking into account the
stochastic-like nature of the samples of all the involved wave
functions for many body problems, a statistical optimization is made
by considering the covariance matrix of these samples. An
eigenvalues system is obtained and solved for the optimal basis set
and by inspecting the rapidly decreasing eigenvalues one may
seriously reduce the necessary number of vectors that insures an
imposed precision. This leads to a potentially significant
improvement in the speed of the SCF calculations and accuracy, as
the statistical properties of a large number of wave functions in an
large spatial domain may be considered. The eigenvalue problem has
to be solved only few times, so that the amount of time added may be
much smaller that the overall iterating SCF calculations.

A simple implementation of the method is presented for a situation
where the analytical solution is known, and the results are
encouraging.\end{abstract}

\maketitle

\section{Introduction}
\label{}

\textit{Ab initio} methods become more and more efficient in various
scientific and technical applications, as the involved physical
principles, numerical methods and computer hardware display a
constantly improvement. The whole effort is sustained by the big
promises of these methods in the general field of the computer
simulation of matter properties, with applications in physics,
chemistry, biology, and many interdisciplinary scientific
researches. However, there are still many problems that demand
better solutions at any level of the \textit{ab initio} methods and
their solving is limited by the enormous computational effort
implied by these kind of calculations even for the very small
clusters of atoms that can be dealt today. As the large
accessibility of supercomputers will be probably delayed for an
unknown period and since there is a continuous need for larger
atomic systems calculations, some progress is certainly needed in
both physical principles (more accurate and elaborate models) and
numerical methods (more precise and fast algorithms) to achieve this
goal.

The physical part of this scenario has already an eight decade
history, starting with the birth of the quantum mechanics. The
Hartree method of the self consistent field (SCF) founded in the
third decade of the last century was soon amended to include the
exchange integral leading to the Hartree-Fock (HF) equations which
still remain the least empirical way for \textit{ab initio}
calculations. Their well known expressions reveal the necessity of
an iterating process, as each wave function depend on the others and
they are present both under the differentiation and integration
operators. Considering the Born Oppenheimer approximation, the left
side of the equation for each particle is composed by the one
electron term, the coulombian interaction between electrons and the
exchange term due to spin:

\begin{eqnarray}
\left[ { - \frac{1}{2}\nabla ^2  + U\left( r \right)} \right]\Psi _i
\left( {\bf{r}} \right) + \sum\limits_j^{} {\int {d{\bf{r}}{\rm{'}}}
\frac{{\left| {\Psi _j \left( {{\bf{r}}{\rm{'}}} \right)} \right|^2
}}{{\left| {{\bf{r}}{\rm{ - }}{\bf{r}}{\rm{'}}} \right|}}\Psi _i
\left( {\bf{r}} \right)} \cr - \sum\limits_j^{} {\delta _{s_i ,s_j}
\int {d{\bf{r'}}} }  \frac{{\Psi _j^* \left( {{\bf{r}}{\rm{'}}}
\right)\Psi _i  \left( {{\bf{r}}{\rm{'}}} \right)}}{{\left|
{{\bf{r}}{\rm{ - }}  {\bf{r}}{\rm{'}}} \right|}}\Psi _j \left(
{\bf{r}} \right) = \varepsilon _i  {\kern 1pt} \Psi _i \left(
{\bf{r}} \right)
\end{eqnarray}
where $\Psi _i \left( {\rm{r}} \right)$ is the one electron wave
function, $U\left( r \right)$ is the potential energy in the nucleus
field, ${\bf{r}}$ and ${\bf{r}}'$ are the vectors to the nucleus of
the two electrons interacting and $s_i $ is the spin index. As
usually, for the sake of simplicity, natural system units have been
used in this equation

About three decades have been necessary for providing a solid
theoretical base, in the framework of the Kohn-Sham (KS) formalism
\cite{KS} of the Density Functional Theory (DFT) , for including the
correlation term, and other decades for its satisfactory evaluation.
Using the properties of an homogenous electron gas and introducing
the functional of density, for a sufficiently slow varying density
$n\left( {\bf{r}} \right) = \sum\limits_{i = 1}^N {\left| {\Psi _i
\left( {\bf{r}} \right)} \right|^2 } $ , the system of equations
reads \cite{KS}:
\begin{equation}
\left\{ { - \frac{1}{2}\nabla ^2  + \varphi \left( {\bf{r}} \right)
+ \mu _c \left( {\bf{r}} \right)} \right\}\Psi _i \left( {\bf{r}}
\right) - \int {d{\rm{r}}'\frac{{n_1 \left(
{{\bf{r}}{\rm{,}}{\bf{r}}{\rm{'}}} \right)}}{{\left| {{\bf{r}}{\rm{
- }}{\bf{r}}{\rm{'}}} \right|}}\Psi _i \left( {{\bf{r}}{\rm{'}}}
\right)}  = \varepsilon _i {\kern 1pt} \Psi _i \left( {\bf{r}}
\right)
\end{equation}
where $\varphi \left( {\bf{r}} \right)$ includes the one and two
electrons potential, $\mu _c \left( {\bf{r}} \right)$ includes the
correlation effects and the last term of the left side of the
equation is the new form of the exchange correction. Although
theoretically more accurate than HF, the KS method has some
necessary simplifying assumptions of the new parts of the model so
that it is often considered slightly empirical in these aspects. For
the most situations this formalism is highly efficient and the
theoretical improved accuracy is present in many of the
applications. However, probably due to its empirical part, it still
fails sometimes, as HF also does in other situations due to the
neglecting of the correlations.

Finally, it is worth to mention the existence of many other more or
less empirical methods, with a greater speed, but with a more
limited scope, with codes commercially available and widely used in
various applications. As the physical model is still a rather simple
one, refining the model is not expected to improve the speed of the
\textit{ab initio} calculations, but rather their accuracy.

Many other post HF models have appeared in the last decades,
offering a multitude of choices for various aspects of the
\textit{ab initio} calculus. Thus, the exchange term may have
several forms, as:  exact HF exchange, Slater local exchange
functional \cite{Slater}, Becke's 1988 non-local gradient correction
to exchange \cite{Becke}, Perdew-Wang 1991 generalized gradient
approximation non-local exchange \cite{PerdWang},\cite{Perd}. For
the correlation term the most accurate expressions seems to be:
Vosko-Wilk-Nusair (VWN) local correlation functional \cite{Vosko},
Perdew and Zunger's 1981 local correlation functional \cite{PerZun},
Lee-Yang-Parr non-local correlation functional \cite{LeeYan},
Perdew-Wang 1991 local correlation functional \cite{PerdWang}.

Concerning the numerical methods, the algorithms and the mathematics
used in the \textit{ab initio} calculations, there are many popular
techniques that may be considered, each of them with their pros and
cons \cite{Press}. Here the possibilities are more diversified, as
the form of the self consistent equations may be purely differential
or integro-differential, due to the exchange and correlation terms.

The differential form is often treated by the Numerov's fifth order
method, which is robust and accurate but is not self starting and
require some initial iterations, as many other point by point
methods of high order. A notable exception should be the forth order
Runge-Kutta method but it is not well suited for boundary conditions
equations as the HF ones. Some shooting method must accompany the
point by point methods and, although this provides the eigenvalue of
the equation (which sometimes is the main goal), it implies an
iterating process that leads to a huge amount of computing effort.
Furthermore, it must be used both for the wave function equation and
for Poisson equation for finding the Hartree-Fock potential
generated by the charge density.

An important step for improving the \textit{ab initio} methods'
efficiency was made by Roothaan \cite{Roth} who transferred the
calculations to linear algebra in the form of a generalized
eigenvalue problem, using non orthogonal basis set. The HF equations
reduce to the following matrix equation which is more suitable for a
numerical calculations:
           \[{\bf{FC = SC\varepsilon }}\]
where ${\bf{F}}$ is the Fock matrix, ${\bf{C}}$ is a matrix of
coefficients of the two electrons interactions, ${\bf{S}}$ is the
overlap matrix of the basis functions, and ${\bf{\varepsilon }}$ is
the matrix of orbital energies. Thus, a class of new numerical
techniques became eligible and an increasing interest for
appropriate choosing of the basis set emerged.

Among the various methods with a reduced need of iterations that are
currently used there are: finite difference method, finite element
method, Galerkin method, collocation method, etc. The most promising
class of methods for such boundary conditions equations is
considered to be the class of spectral and pseudospectral  methods
\cite{Boyd}, \cite{Freis}, as they have already been successfully
used in various other fields. Their evanescence property
(exponential decay of the error with the number of sampling points)
and the absence of iteration processes are very attractive, but the
main problem is the bad conditioning that often appears in the
linear algebra implied. However, the matrix conditioning numbers are
generally satisfactory if a low dimension of the vectorial space is
chosen, but this tends to increase the errors if the basis set is
not an optimal one.

We consider that the spectral methods in the \textit{ab initio}
calculations have not been entirely exploited, as even the basis
sets seem to be still "empirically" chosen: by subjective
considerations (as Chebyshev polynomials are the usual basis of
choice \cite{Boyd}), or artificial approximations that facilitate
the analytical calculations with a great impact on the speed of the
numerical part (for example the Gaussian basis). As the number of
vectors used by a basis must be always finite and as small as
possible to minimize the matrix condition number and the overall
computational effort, the problem seems to be the choice of an
optimum basis that provide the minimal error when the dimension of
the vectorial space of the wave function representation is highly
reduced.

In the next chapter we propose a method for properly choosing the
basis set for achieving this minimization of the errors when dealing
with a smaller number of vectors that is usually needed. An example
for the implementation of the method will be later presented and
some conclusions and further suggestions will be drawn.

\section{The stochastic nature of the wave functions' samples for many body problems}

The radial wave function in a hydrogenoid atom, which is the primary
natural approximation of the many body wave function has the well
known form \cite{Messiah}:

    \begin{equation}
R_{n\ell}\left( r \right) =\frac{2}{n^2} \sqrt {\frac{{\left( {n -
\ell - 1} \right)!}}{{\left[\left( {n + \ell}\right)!\right]^3}} }
\left( {\frac{{2r}}{n}} \right)^\ell e^{-\frac{r}{n}}L_{n - \ell -
1}^{2\ell + 1} \left( {\frac{{2r}}{n}} \right)
\end{equation}
where $n$
 is the principal quantum number, $\ell$
 is the angular momentum number, $r$ is the distance to the nucleus in Bohr radius and $L_n^\ell \left( r \right)$
 are the associate Laguerre polynomials.

\begin{figure}
\centering
\includegraphics{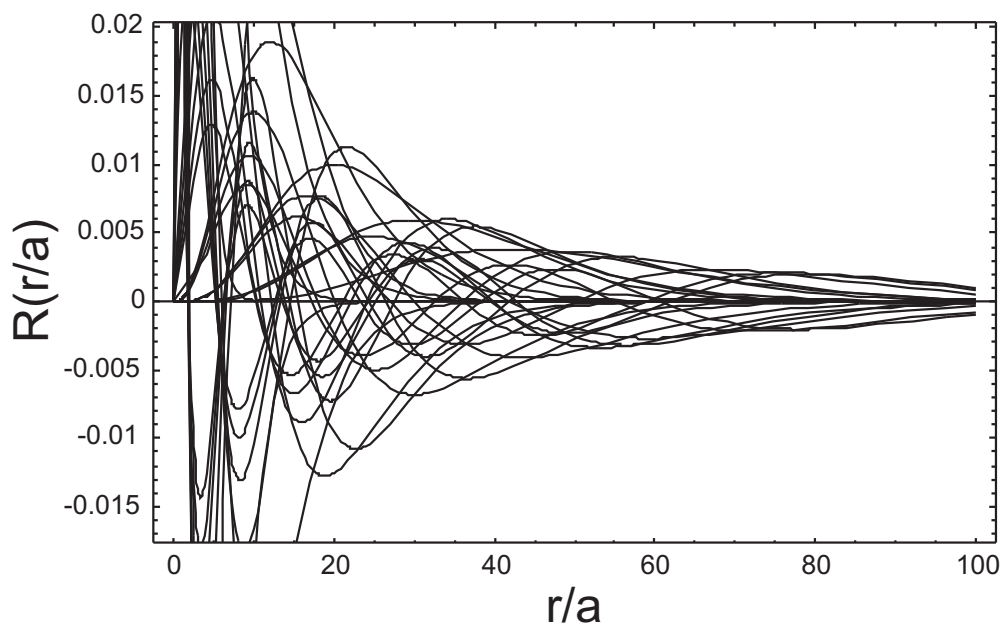}
\caption{The various wave functions in a hydrogen-like atom for
$n=1,2,...,7$} \label{fig1}
\end{figure}

Every electron moving in the potential determined by the charge
distribution created by such type of wave functions is thus
influenced by all the others electrons. For systems with many
electrons, the family of functions has the appearance shown in
figure \ref{fig1} (generated using \textit{Mathematica} software).
The asymptotic behavior of the wave functions allows one to deal
with a finite interval, but this interval seems to be quite large
for the usual methods for point to point differential equations .
Thus, the errors are very important for most of the orbitals that
imply distances above 15-20 Bohr radius, but the influence of the
further regions is clearly present in the real system, and needs to
be considered.

\begin{figure}
\centering
\includegraphics{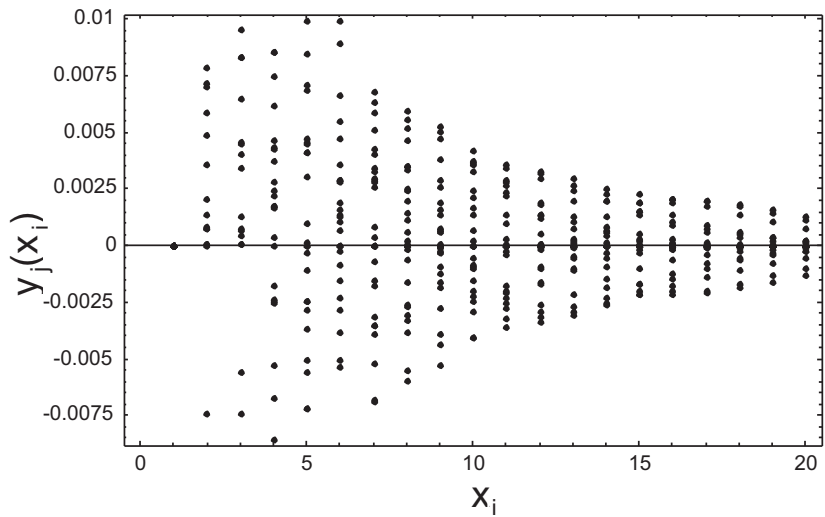}
\caption{The equidistant samples of the wave functions in a
hydrogen-like atom for $n=1,2,...,7$} \label{fig2}
\end{figure}

If one samples the distance to the nucleus, a pseudo-random
distribution of the local values of these wave function appears as
shown in figure \ref{fig2}.

Although the entropy displayed by these values is not very high, it
is possible to consider a stochastic influence between the
electrons, and use the second order statistic's methods for dealing
with them. The concept of the "mean field", with a clear stochastic
connotation is present in many of the physical models accepted
now-a-days but this context has not been considered enough in the
present theories, except for some low order Monte-Carlo methods.

\section{ Using the Karhunen-Loeve theorem for calculating the optimum basis
set}

The spectral methods used for solving the eigenvalue problems in
simple or generalized form, first approximate the solution as a
linear combination of continuous functions - basis vectors- and then
plug in this solution in the original equations to determine the
coefficient matrix.
\begin{equation}y\left( x \right) \approx \sum\limits_{i =
0}^N {c_i \varphi _i \left( x \right)}\label{eq11}
\end{equation}

The domain of the independent variable is sampled in a number of
points $x_j $ (as in the collocation method) where the equations are
supposed to be satisfied, and the resulting linear system is solved
for the coefficients.
\begin{equation}y\left( {x_j } \right) = \sum\limits_{i = 0}^N {c_i \varphi _i
\left( {x_j} \right)} ,\quad j = 0,...,N \label{eq11a}
\end{equation}

As the values of $y\left( x \right)$ are known for the boundary
points, a system of $N + 1$ equations is obtained, and its solution
gives the unknown coefficients. Although it seems to be a very
simple procedure, there are several problems that may occur and must
be somehow taken into account. First of all, if one uses a simple,
equidistant sampling, the convergence of the method, theoretically
exponential, is lost due to the Runge phenomenon. That is why
various non equidistant sampling methods are currently used with a
smaller step at the extremities of the interval (as Chebyshev,
Gauss-Lobatto or Legendre points). The second, and the most serious
problem is the fact that the relation (\ref{eq11}) is still an
approximation and eventually the equality is true for a huge if not
an infinite number of basis vectors, for the most part of the
applications. For practical purposes, only a very limited number of
these basis vectors may be used, for two reasons:

- The computational effort increases with at least $N^3 $, limiting
the number of wave functions that may be dealt with reasonable
timing.

-Increasing $N$ always increases the condition number of the matrix
and hence serious errors occur for $N$ values above 10 - 20.

That is why the basis set must be very carefully chosen, as it may
prove itself as crucial for the performance of the whole calculus.

An important observation must be now taken into account: the
coefficients of the linear combination (\ref{eq11}) carry some
amount of redundancy in all but one case. This unique case implies
that all of them to be uncorrelated, in the second order statistics
meaning:
    \begin{equation}
E\left\langle {c_i c_j } \right\rangle  = \lambda _i \delta _{ij}
,\quad i,j = 0,...,N \label{11}
\end{equation}
where the symbol $E\left\langle . \right\rangle $ stands for the
expected value operator and $\delta _{ij}$ is the Kronecker symbol.

It follows that if the coefficients are uncorrelated, the
information lost by truncating the series (\ref{eq11}) to a number
$M < N$ of terms (for the above presented reasons) is minimized, and
the truncating errors are also minimized.

Hence, a basis set that ensures a pairwise uncorrelated set of
coefficients must be found, which is possible by using the
Karhunen-Lo\`{e}ve theorem \cite{Karh} \cite{Loeve}. It states that
the basis set that ensures the minimization of the errors due to the
truncation of the decomposition expression of a continuous function
is the solution of the integral equation:
    \begin{equation}
\int\limits_D {K_{yy} \left( {x,x'} \right)\varphi _i \left( {x'}
\right)dx'}  = \lambda _i \varphi _i \left( x \right)\label{eq12}
\end{equation}
which is a second kind homogeneous Fredholm equation, i.e. an
eigenvalue problem. Here $D$
 is the orthogonality domain of the eigenvectors $\varphi _i \left( x \right)
$ that form the optimal basis set, $\lambda _i $
 are the associate (positive) eigenvalues and $K_y \left( {x,x'}
 \right)$ is the autocorrelation function of the initial function $y\left( x \right)$
 defined by:
    \begin{equation}
K_{yy} \left( {x,x'} \right) = E\left\langle {y\left( x
\right)y\left( {x'} \right)} \right\rangle
\end{equation}
The set of orthogonal functions obtained from (\ref{eq12}) is
complete and the functions are square-integrable.

If the initial function $y\left( x \right)$ is the wave function ,
one may consider that the eigenvalues represent the probability
density associated with each mode, and according to the KL theorem
the set of basis functions is the optimal from all possible sets,
i.e. the decomposition series converges as rapidly as possible. The
method also minimizes the representation entropy and is equivalent
with the minimization of the mean-square error resulted from the
truncation.

It is widely applied in signal theory (detection, estimation,
pattern recognition, noise rejection, data compression for storage
and image processing), physics (stochastic turbulence processes
\cite{Lumley}) and biology \cite{Brooks}, under various names:
Uncorrelated Coefficients Series, Principal Component
Analysis\cite{Hotelling}, Hotelling Analysis \cite{Hotelling},
Quasiharmonic Modes \cite{Brooks}, Proper Orthogonal Decomposition
\cite{Lumley}, Singular Value Decomposition \cite{Golub} etc.

The stochastic-like nature of the samples of all the wave function
in a many body problem, suggests that this method could be also
applied to the SCF problems. Since the numerical processing involves
the discretization of the implied functions, the KL transform is
often met in matrix form with finite dimensional vectors.
    Considering a total number of $N_w $
hydrogen-like wave functions and sampling all of them in $N_s $
points, we obtain $N_w $ column matrices each of them containing
$N_s $ elements:

\begin{equation}
{\bf{Y}}_j  = \left( \begin{array}{l}
 y_j \left( {x_1 } \right) \\
 y_j \left( {x_2 } \right) \\
 . \\
 . \\
 . \\
 y_j \left( {x_{N_s } } \right) \\
 \end{array} \right),\quad j = 1,2,...,N_w \quad
\end{equation}

The resulting $N_s  \times N_w $
 matrix
    \begin{equation}
{\bf{Y}} = \left( {{\bf{Y}}_0 \quad {\bf{Y}}_1 \quad .\quad .\quad
.\quad {\bf{Y}}_{N_w } } \right)
\end{equation}
contains all the samples of all the wave function for hydrogen-like
atoms. By averaging the similar samples and subtracting the result
from each sample of each wave function one obtains a matrix with
columns of zero centered values, as the KL transform demands.
\begin{equation}
{\bf{Y}}^c  = \left( {{\bf{Y}}_1^c \quad {\bf{Y}}_2^c \quad .\quad
.\quad .\quad {\bf{Y}}_{N_w }^c } \right)
\end{equation}

with
    \begin{equation}
{\left(\bf{Y^c}\right)}_{ij}  = y_j \left( {x_i } \right) -
\frac{1}{{N_w }}\sum\limits_{j = 1}^{N_w } {y_j \left( {x_i }
\right)} ;\;i = 0,1,...,N_s ;\;j = 0,1,...,N_w
\end{equation}

Using the centered samples matrix ${\bf{Y}}^c $
 and its transpose $\left[ {{\bf{Y}}^c } \right]^T $
 one may construct the $N_s  \times N_s $
covariance matrix defined by
    \begin{equation}
    {\bf{K}}_{yy}  = E\left\langle {{\bf{Y}}^c \left[ {{\bf{Y}}^c } \right]^T } \right\rangle
    \label{eq37}
    \end{equation}

Using these matrices and sampling the independent variables $x$
 and $x'$
in eq. (\ref{eq12})  it is transformed in an eigenvalue system for
the covariance matrix:
    \begin{equation}
{\bf{K}}_{yy} \bf{\Phi } _j  = \lambda _j \bf{\Phi } _j ,\quad j =
1,...,N_s \label{eq38}
\end{equation}
which has to be solved to obtain the column matrices $\Phi _j $
 containing the samples of the needed basis function $\varphi _i (x)$
, $i = 0,1,...,N$ . One should take sufficient number samples to
assure $N_s \ge N + 1$ , where $N$
 is the dimension of the primary space of the decomposition.
        The eigenvalues $\lambda _j $
obtained by solving the  covariance matrix are supposed to be
strongly decreasing with $j$ , and this was numerically checked, as
shown in the next chapter. The first $M$ largest eigenvalues
indicate the corresponding eigenvectors $\varphi _i $ , with $i =
0,1,...,M$ that should be taken into account in a truncated
decomposition with acceptable errors. The more rapidly the
eigenvalues decrease, the smaller is the number of vectors in the
basis set that must be taken into account. In the next chapter we
will show a simple example of implementation of this technique, and
some preliminary results for the covariance matrix and the basis
set.

Mathematically, the KL equation (\ref{eq12}) is equivalent to a
transformation which diagonalizes a given matrix K and turns it to a
canonical form
    \begin{equation}
{\bf{K}}_{yy} {\rm{ = }}{\bf{U\Lambda V}}
\end{equation}
where ${\bf{\Lambda }}$ is a diagonal matrix. Indeed, by forming a
matrix $\bf{\Phi } $ , who's columns are the basis vectors $\bf{\Phi
} _j $

    \begin{equation}
\bf{\Phi }  = \left( {\bf{\Phi } _1 \quad \bf{\Phi } _2 \quad .\quad
.\quad .\quad \bf{\Phi } _{N_w } } \right)
\end{equation}
we may define a transformation from the primary space of the samples
$ y_j \left( {x_i } \right)$
 to a secondary space as:
    \begin{equation}{\bf{Z}} = \bf{\Phi } ^T {\bf{Y}}\label{eq310}\end{equation}

In this space the covariance matrix defined similarly with eq.
(\ref{eq37}) has a diagonal form:
    \begin{equation}
\begin{array}{l}
 {\bf{K}}_{ZZ}  = E\left\langle {{\bf{ZZ}}^T } \right\rangle  = E\left\langle {{\bf{\Phi }}^T {\bf{Y}}\left( {{\bf{\Phi }}^T {\bf{Y}}} \right)^T } \right\rangle  = {\bf{\Phi }}^T E\left\langle {{\bf{YY}}^T } \right\rangle {\bf{\Phi }} \\
  = {\bf{\Phi }}^T {\bf{K}}_{YY} {\bf{\Phi }} = $Diag$\left( {\lambda _i } \right) \\
 \end{array}
\end{equation}
where we took into account eqs. (\ref{eq38}), (\ref{eq310}) and the
orthogonality of the basis set ${\bf{\Phi }}^T {\bf{\Phi }} =
{\bf{I}}$. It then makes possible to use one of the many well
developed methods for diagonalizing the covariance matrix in order
to find the eigenvectors and to select those of them which
correspond to the largest eigenvalues as the optimal basis set.

\section{Example of implementation of the method and some results}

In order to test the above described procedure we considered a
simple situation for the ground state hydrogen atom which has an
analytical solution for the wave function and a precisely known
orbital energy. This allows an objective test by comparing our
results with an exact one and is also a rather difficult situation,
because the covariance matrix is constructed using all the 28
different orbitals for $n=1-7$ and $\ell=0-6$. Thus, the 1S orbital
is at the extremity of the whole range of wave functions and better
results may be expected for an intermediate orbital. The radial part
of the wave function may be obtained analytically by imposing the
boundary conditions $y(x)=y_i=0$ for $x=0$ and $y(x)=y_f$ for $x=b$
(in Bohr radius) to the radial part of the Schr\"{o}dinger equation
\begin{equation}
 - \frac{1}{2}y''\left( x \right) + \left[ { - \frac{Z}{{x + \varepsilon }}  + \frac{{\ell(\ell + 1)}}{{2x^2 + \varepsilon }}} \right]y\left( x \right) =
 E_n y\left( x \right)
\label{41}
\end{equation}
where $x=r/a_0$, $y(r)=r R(r)$, $a_0$ is the Bohr radius and, in
order to avoid the singularity in the origin in the numerical
calculations, we added the very small constant $\epsilon$ (say
$10^{-10}$) at the denominator of the coulombian and orbital terms
(that does not affect seriously the numerical results).

For K-shell electrons and $Z=1$, the solution in a the region
$(0,b)$ with $b$ finite, may be expressed in terms of exponential
integral functions $\textrm{Ei}\left(z\right)$ as the symbolic
software generated expression:

\begin{equation}
y\left( x \right) = \frac{{e^{b - x} \left\{ {e^{2\left( {x +
\varepsilon } \right)} \varepsilon  - \left( {x + \varepsilon }
\right)\left[ {e^{2\varepsilon }  - 2\varepsilon
\textrm{\textrm{Ei}}\left( {2\varepsilon } \right) + 2\varepsilon
\textrm{Ei} \left( {2x + 2\varepsilon } \right)} \right]}
\right\}}}{{e^{2\left( {b + \varepsilon } \right)} \varepsilon  +
2\varepsilon \left( {b + \varepsilon } \right)\textrm{Ei}\left(
{2\varepsilon } \right) - \left( {b + \varepsilon } \right)\left[
{e^{2\varepsilon }  + 2\varepsilon \textrm{Ei}\left( {2b +
2\varepsilon } \right)} \right]}}y_f
\end{equation}

We used the natural units and thus the energy is expressed in
Hartree ($E_n=-0.5$ Hartree for $Z=1,n=1,\ell=0$). The boundary
condition $y_f$, appearing as a global coefficient of the solution
may be taken arbitrary or may be determined from the normalization
of the resulted wave function.

\begin{figure}
\centering
\includegraphics{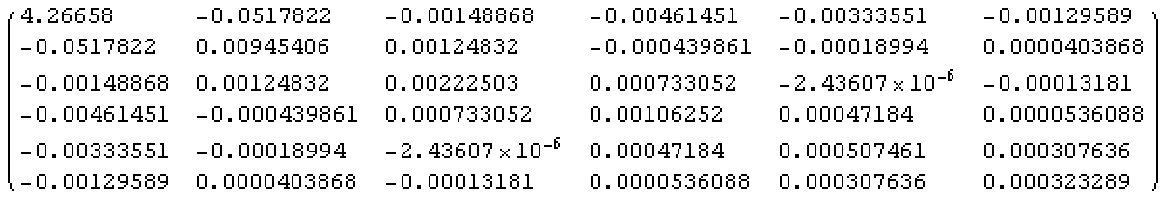}
\caption{An example of the obtained covariance matrix $\bf{K}_{yy}$
for $N_s=6$ samples and $N_w=28$ wave functions} \label{fig3}
\end{figure}

Using a parameter $b=40$ Bohr radius and the samples of the 28 wave
function we construct the covariance matrix according to eq.
(\ref{eq37}). We exemplify this matrix only for 6 equidistant
samples for each wave function, for space saving in figure
\ref{fig3}.

\begin{figure}
\centering
\includegraphics{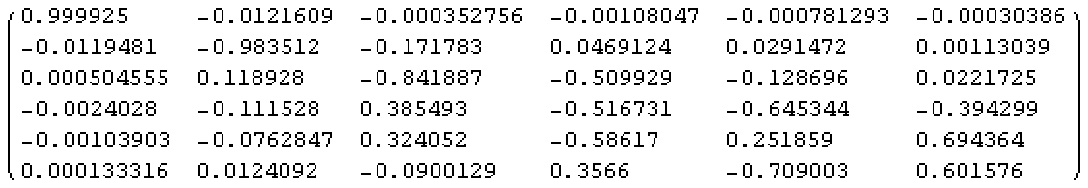}
\caption{An example of the KL transform matrix $\bf{\Phi}$, for
$N_s=6$ samples and $N_w=28$ wave functions. Each column contains
the samples of a function $\bf{\Phi}_i$ of the optimal basis set}
\label{fig4}
\end{figure}

By solving numerically the eigenvalue problem for this covariance
matrix we obtain the KL transformation matrix $\bf{\Phi}$ presented
in figure \ref{fig4}, also only for 6 samples per wave function.

\begin{figure}
\centering
\includegraphics{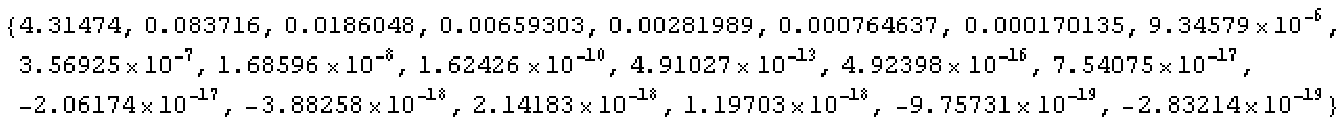}
\caption{An example of the eigenvalues $\lambda_i$ obtained for
$N_s=20$ samples and $N_w=28$ wave functions} \label{fig5}
\end{figure}

For a realistic calculation, we used $N_s=20$ equidistant samples
and from the covariance matrix (not shown here) we obtained the
eigenvalues listed in figure \ref{fig5}. The equidistant sampling
method was chosen for simplicity of this exemplification but
increasing errors are expected at the extremities of the interval
due to the Runge phenomenon. Of course, in practice we recommend
more appropriate sampling methods like those mentioned in chapter 3.

One may see that the eigenvalues are indeed rapidly decreasing,
confirming the possibility of reducing the dimension of the basis
set, as it was anticipated in the previous subsection. It follows
from the presented values that only 8-10 eigenvectors should be
considered, the others corresponding to much smaller eigenvalues.

\begin{figure}
\centering
\includegraphics{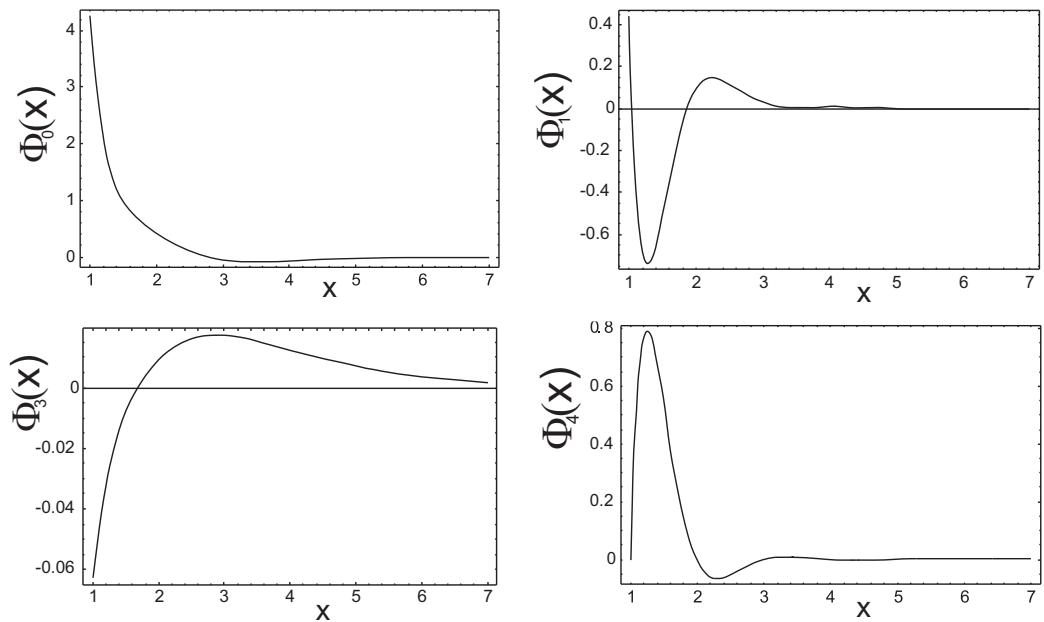}
\caption{Some of the basis set functions obtained for $N_s=20$
samples and $N_w=28$ wave functions} \label{fig6}
\end{figure}

For solving the differential equations using spectral methods, one
needs the derivatives of the basis function. Taking into account
that these function are known in a small number of points (their
samples) it is not effective to use a direct numerical
differentiation. Instead, one may interpolate those samples for each
of the basis functions and thus obtain an analytical expression, for
example a polynomial. For the sake of simplicity, we used a Lagrange
interpolation (but some other methods may be investigated for better
results) and obtained the basis functions as exemplified in figure
\ref{fig6}.

\begin{figure}
\centering
\includegraphics{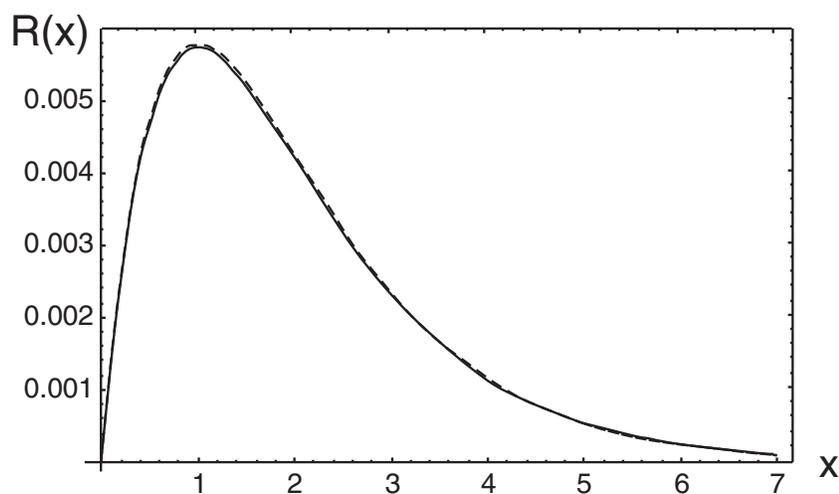}
\caption{A comparison of the analytical solution (dashed) with the
numerical one (continous) obtained for $N_s=20$ samples and $N_w=28$
wave functions} \label{fig7}
\end{figure}

Using these analytical form for the basis set, we checked the
numerical solving of the radial differential equation \ref{41} with
the following parameters: $\ell=0,\;n=1, \;E_n=-0.5,\; N_s=8,
\;a=0,\; b=7,\; y_f=10^{-4},\;\epsilon=10^{-10}$. We obtained a wave
function as presented in figure \ref{fig7}, which is very close to
the analytical one presented dashed in the same figure.

\begin{figure}
\centering
\includegraphics{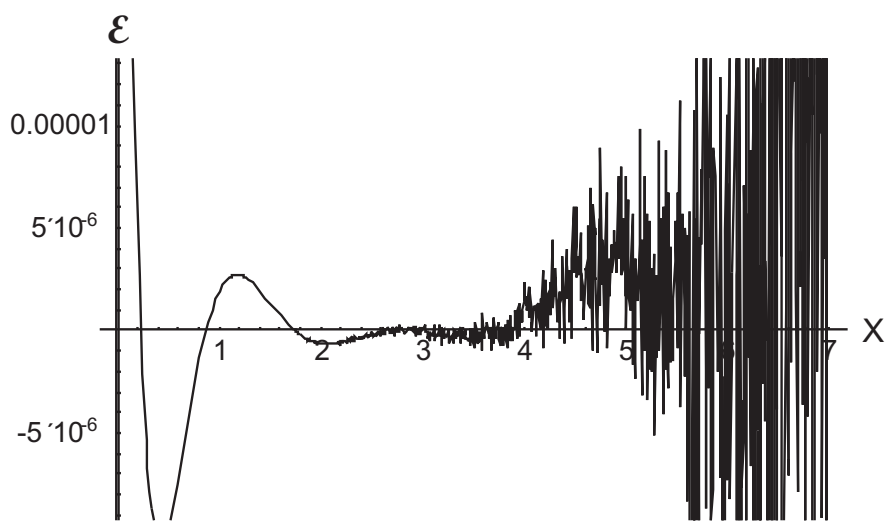}
\caption{Difference between the two members of the radial equation
obtained for $N_s=20$ samples and $N_w=28$ wave functions}
\label{fig8}
\end{figure}

The errors introduced by the interpolated basis set, obtained as the
difference of the two members of the differential equation are
presented in figure \ref{fig8}, where one may see the effect of the
uniform sampling that we used for this example: the errors are
highly increasing towards the boundaries. Also, the result of the
bad conditioning of the system (\ref{eq11a}) may be noticed for high
values of x both in figure \ref{fig7} and figure \ref{fig8}. It is
possible that this conditioning should be improved if proper
sampling and interpolation techniques are used. Anyway, the result
is very encouraging, taking into account that for other basis sets
(for example gaussian), in similar conditions, clearly higher errors
are displayed.

\section{Conclusions}

Although the above presented method is highly susceptible for
improvements, the simple example presented for implementing the
concept of the optimal basis set obtained using the covariance
matrix over the set of typical wave functions gives satisfactory
results. The most important feature of the method is the virtual
increase in speed due to the theoretically smallest number of basis
functions needed and their construction using the information
contained in a big number of possible orbitals.

The price paid is the necessity to calculate and diagonalize the
covariance matrix, but it is important to notice that it must be
done \textit{only at the start} of the iteration process and maybe
in some intermediate steps. The assertion above is experimentally
proved by the fact that in data processing an initial set of samples
may be used for constructing covariance matrices adequate for other
different situations.

Improvements of the presented example may be made in the sampling
and interpolations processes and the principle may be also applied
to other various spectral methods used by \textit{ab initio}
calculations.

\ack This work was supported by The Council of Scientific Research
in Higher Studies in Romania (CNCSIS) under Grant 556/2008.

--------------------------------------------------------------------------------------------

\end{document}